\documentclass[twocolumn]{jpsj2} %% two-column layout
\title{Magnetic and Structural Studies of the Quasi-Two-Dimensional Spin-Gap System (CuCl)LaNb$_2$O$_7$ }

\author{Makoto \textsc{Yoshida}$^{1}$\thanks{E-mail address: yopida@issp.u-tokyo.ac.jp}, 
Nobuyuki \textsc{Ogata}$^{1}$, 
Masashi \textsc{Takigawa}$^{1}$\thanks{E-mail address: masashi@issp.u-tokyo.ac.jp}, 
Jun-ichi \textsc{Yamaura}$^{1}$, 
Masaki \textsc{Ichihara}$^{1}$, 
Taro \textsc{Kitano}$^{2}$, 
Hiroshi \textsc{Kageyama}$^{2}$, 
Yoshitami \textsc{Ajiro}$^{2}$, 
and Kazuyoshi \textsc{Yoshimura}$^{2}$}

\inst{$^{1}$Institute for Solid State Physics, University of Tokyo, Kashiwa, Chiba 277-8581 \\ 
$^{2}$Department of Chemistry, Graduate School of Science, Kyoto University, Kyoto 606-8502}

\abst{We report magnetization, nuclear magnetic resonance (NMR), nuclear quadrupole resonance (NQR), and 
transmission electron microscopy (TEM) studies on the quasi-two-dimensional spin-gap system 
(CuCl)LaNb$_2$O$_7$, a possible candidate for the $J_1$-$J_2$ model on a square lattice. 
A sharp single NQR line is observed at the Cu and Cl sites, indicating that 
both Cu and Cl atoms occupy a unique site. However, the electric field gradient tensors at the Cu, Cl, and La sites 
do not have axial symmetry. This is incompatible with the reported crystal structure. Thus the 
$J_1$-$J_2$ model has to be modified. 
We propose alternative two-dimensional dimer models based on the NMR, NQR, and TEM results. 
The value of the hyperfine coupling constant at the Cu sites indicates that the spin density 
is mainly on the $d(3z^2-r^2)$ orbital ($z \parallel c$). At 1.5 K, Cu- and Nb-NMR signals disappear above the critical field 
$B_{c1} \simeq   $ 10.3 T determined from the onset of the magnetization, 
indicating a field-induced magnetic phase 
transition at $B_{c1}$.}

\kword{NMR, NQR, TEM, (CuCl)LaNb$_2$O$_7$, spin-gap, quantum spin system}

\begin{document}
\maketitle

\section{Introduction} %% No sections necessary for express letters, letters and short notes 
\begin{figure}[tb]
\begin{center}
\includegraphics[width=7.5cm]{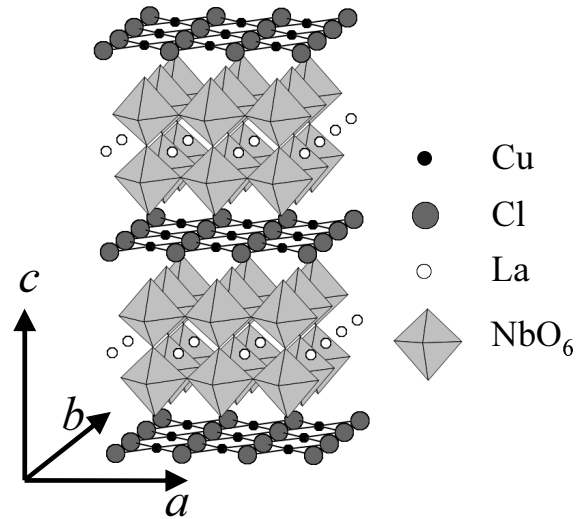}
\end{center}
\caption{Schematic drawing of the crystal structure of (CuCl)LaNb$_2$O$_7$.\cite{c1}}
\label{structure}
\end{figure}

Low-dimensional quantum spin systems with frustrating interactions have recently been studied 
intensively both from experimental and theoretical aspects. The $S$ = 1/2 antiferromagnet on a square lattice  
with the nearest-neighbor and the next-nearest-neighbor interactions (the $J_1$-$J_2$ model) is a 
well known example. 
When the nearest neighbor interaction $J_1$ is antiferromagnetic ($J_1 > 0$), the ground state has a 
N\'{e}el order with the wave vector ($\pi , \pi $) for $J_2$/$J_1$ $< $ 0.38 while a collinear order with the wave 
vector ($\pi , 0$) occurs for $J_2$/$J_1$ $>$ 0.52.\cite{a1,a2,a3} A spin liquid phase is proposed 
between these phases near $J_2$/$J_1$ $\simeq $ 0.5.\cite{a1,a2,a3} A spin liquid phase is also proposed 
for ferromagnetic $J_1$ ($J_1 < 0$) between the collinear phase for $J_2$/$J_1$ $<$ $-$0.51 
and the ferromagnetic phase for larger $J_2$/$J_1$.\cite{a3} 
The series of compounds (Cu$X$)LaNb$_2$O$_7$ ($X$ = Cl, Br) 
is a candidate for the $J_1$-$J_2$ model with mixed 
ferro- and antiferromagnetic interactions.\cite{b1,b2,b3} 
The structure is reported to be tetragonal (space group $P4/mmm$) with the  
Cu$X$ square lattice\cite{c1,c2} as shown in Fig.~\ref{structure}.
In fact, (CuCl)LaNb$_2$O$_7$ has a singlet ground state with a spin-gap\cite{b1,b2} while 
(CuBr)LaNb$_2$O$_7$ shows a collinear antiferromagnetic order.\cite{b3} 

In (CuCl)LaNb$_2$O$_7$, the spin-gap is estimated to be 27 K by fitting 
the susceptibility data to the isolated dimer model.\cite{b1} 
The neutron inelastic scattering results also exhibit magnetic 
excitations with the gap of 2.3 meV, although the resolution is limited.\cite{b1} 
On the other hand, the high-field magnetization measurements revealed a sudden increase 
at the critical field $B_{c1}$ = 10.3 T.\cite{b2} 
The Zeeman energy at $B_{c1}$, $g\mu_{\mathrm{B}} B_{c1}/k_{\mathrm{B}}$ = 15 K, is apparently not sufficient to 
close the spin-gap at zero field. 
In addition, the magnetization saturates at $B_{c2}$ = 30.1 T.\cite{b2} 
The large interval between $B_{c1}$ and $B_{c2}$ indicates a large width of the triplet dispersion. 
This is inconsistent with the neutron inelastic scattering data indicating that the triplet 
mode is almost dispersionless.\cite{b1} 

There are also open issues about the crystal structure. 
The neutron diffraction study indicates that the Cl position deviates from the ideal 
(0, 0, 1/2) position with the $C_4$-symmetry in the tetragonal space group $P4/mmm$.\cite{c2} 
If the structure deviates from the ideal square lattice, 
various ways to form a spin-gap become possible.\cite{c3} Unambiguous determination of the structure 
is crucial for good understanding of a frustrated quantum spin system. 

In this paper, we report magnetization, 
nuclear magnetic resonance (NMR), nuclear quadrupole resonance (NQR), and 
transmission electron microscopy (TEM) studies on (CuCl)LaNb$_2$O$_7$. 
Our results indicate that none of the Cu, Cl, and La sites has the $C_4$-symmetry, incompatible with 
the reported tetragonal structure. 
Thus the $J_1$-$J_2$ model for the simple square lattice has to be modified. 
We propose two-dimensional dimer models which are consistent with our NMR, NQR, and TEM results. 
Cu- and Nb-NMR signals disappear above the critical field $B_{c1} \simeq  $ 10.3 T at low temperatures, 
indicating a field-induced magnetic phase transition at $B_{c1}$.

\section{Experiment} 
\begin{figure}[tb]
\begin{center}
\includegraphics[width=7.5cm]{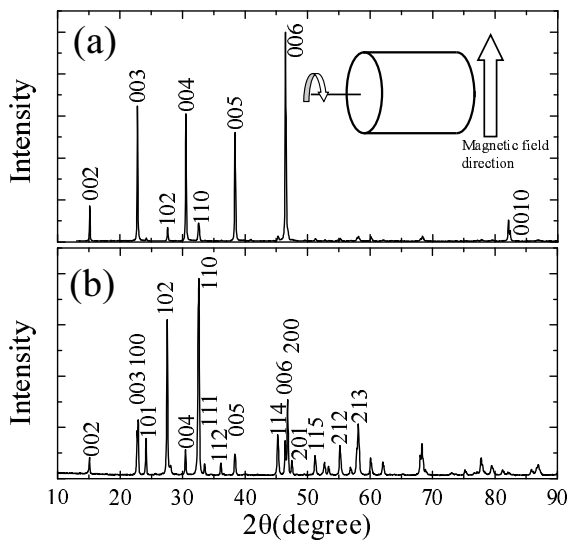}
\end{center}
\caption{X-ray diffraction patterns of the (a) aligned and (b) unaligned powder samples. 
The inset illustrates the method for preparing the aligned sample.}
\label{e1}
\end{figure} 

The powder sample of (CuCl)LaNb$_2$O$_7$ was synthesized by the ion-exchange reaction 
of RbLaNb$_2$O$_7$ with CuCl$_2$.\cite{b1,c1} 
The magnetically aligned sample was prepared by rotating a cylindrical tube filled with 
the (CuCl)LaNb$_2$O$_7$ powder and epoxy (STYCAST1266) in a magnetic field $B$ = 7 T. 
The magnetic field was applied perpendicular to the rotating axis as shown in the inset of 
Fig.~\ref{e1}. 
In this configuration, the magnetic hard axis tends to align along the rotating axis.  
The X-ray diffraction pattern for the aligned sample was obtained with a MacScience M03X 
diffractometer using a monochromatic Cu-$K \alpha$ radiation as shown in Fig.~\ref{e1} (a). 
Here the aligned axis was kept perpendicular to the scattering wave vector. 
Compared with the diffraction pattern for the unaligned powder shown in Fig.~\ref{e1} (b), 
the peaks corresponding to (00$n$) planes are strongly enhanced in the aligned sample, where $n$ is integer. 
This indicates that the $c$-axis is the magnetic hard axis. 
The rocking curve for these peaks with the half width of about 1.5$^\circ $ shows successful alignment. 
However, other peaks such as (102) or (110) are also observed in the aligned sample, indicating that a 
certain fraction of the sample remains unaligned. 

The magnetic susceptibility $\chi $ was measured for the unaligned powder sample with a SQUID magnetometer 
(Quantum Design, MPMS). 
NQR measurements in zero magnetic field have been performed for the $^{63}$Cu and 
$^{35}$Cl nuclei in the unaligned powder sample. Nuclear spin-lattice relaxation rates were measured 
by the inversion recovery method. NMR measurements in magnetic fields up to 11.7T 
have been performed on the aligned sample 
with the magnetic field applied along the $c$-axis for 
$^{63, 65}$Cu, $^{35, 37}$Cl, $^{93}$Nb, and $^{139}$La nuclei. 
The TEM experiments were carried out at room temperature using a JEM2010F system 
with an operating voltage of 200 kV. The specimen was finely ground in methanol and then 
placed on a Cu microgrid mesh for TEM observation. 

\section{Results and Analysis} 

We first analyze the temperature dependence of $\chi $ in section 3.1. 
In section 3.2, we show the NQR results, which demonstrate that 
both Cu and Cl atoms occupy a unique site and there is no apparent disorder. 
The electric field gradient tensors at the Cu, Cl, and La sites are investigated from the field swept 
NMR spectra in section 3.3. These results indicate deviation of the crystal structure from 
the reported one with a simple square lattice. In section 3.4, the temperature dependences 
of the magnetic hyperfine shifts at the Cu and Cl sites are determined. 
Then the value of spin-gap is estimated. The hyperfine coupling constants 
are also determined at the Cu, Cl, and Nb sites. In section 3.5, the high 
field behavior above $B_{c1}$ is investigated at the Cu and Nb sites. 
Finally, we show the TEM result in order to provide additional 
information about the structure. 

\subsection{Magnetic susceptibility measurements} 

\begin{figure}[tb]
\begin{center}
\includegraphics[width=7.5cm]{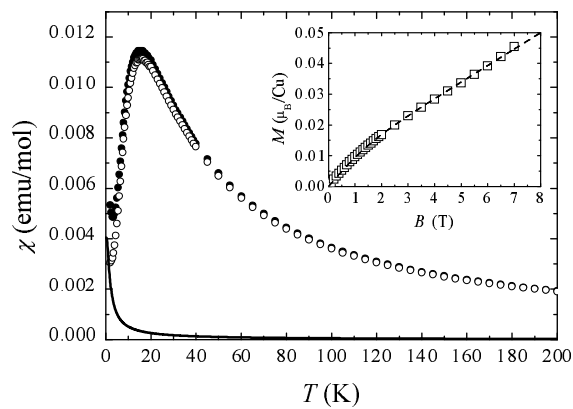}
\end{center}
\caption{Temperature dependence of $\chi $ measured for the unaligned powder sample at 1 T. 
The solid circles represent the raw data. 
The open circles represent $\chi $ after subtracting the impurity contribution 
shown by the solid line. The inset shows the magnetic field ($B$) dependence of the magnetization at 2 K. 
The dashed line is the fit to the form $M = NgS\mu _{\mathrm{B}}B_S(x) + AB$.}
\label{kai}
\end{figure} 

Figure~\ref{kai} shows the temperature ($T$) dependence of $\chi $ measured at the field of 1 T. 
The peak at 16 K and rapid decrease at lower temperatures are quite consistent with the previous report,\cite{b1} 
indicating a singlet ground state. Below 4 K, however, $\chi $ increases with decreasing 
temperature probably due to magnetic impurities. In order to investigate the impurity contribution, 
we also measured magnetization $M$ up to 7 T. The inset of Fig.~\ref{kai} 
shows the $B$-dependence of $M$ at 2 K. 
The data can be fitted by the sum of a term proportional to the Brillouin function 
$B_S(x)$ and a $B$-linear component $M = NgSB_S(x) + AB$ ($x = gS\mu _{\mathrm{B}}B/k_{\mathrm{B}}T$). 
A good agreement was obtained for the values $S$ = 1/2, $N$ = 0.72 \%/Cu, and 
$N_{\mathrm{A}} \mu _{\mathrm{B}}A$ = 3.0 $\times $ 10$^{-3}$ emu/mol ($N_{\mathrm{A}}$ is the Avogadro constant) 
as shown by the dashed line in the inset of Fig.~\ref{kai}. 
The first term leads to the free spin contribution to $\chi $, $N_{\mathrm{A}}NgS\mu _{\mathrm{B}}B_S(x)/B$, 
due to isolated magnetic impurities (the solid line in Fig.~\ref{kai}). 
By subtracting this term from the measured $\chi $, we obtained the corrected $\chi $ 
shown by the open circles. 
The corrected $\chi $ maintains a finite value (3$\times 10^{-3}$ emu/mol) as $T$  goes to 0. 
It is also detected in the magnetization curve as the $B$-linear component. 
This is much larger than 
the typical value of the Van Vleck susceptibility of Cu$^{2+}$ ($10^{-5} \sim 10^{-4}$ emu/mol). 
The NMR shifts at all sites 
do not show such large residual values at low temperatures as we discuss later. 
Hence the large residual susceptibility at low $T$ is considered to be extrinsic, e.g. due to 
a secondary phase.

\subsection{NQR spectra and relaxation rates}
\begin{figure}[tb]
\begin{center}
\includegraphics[width=7.5cm]{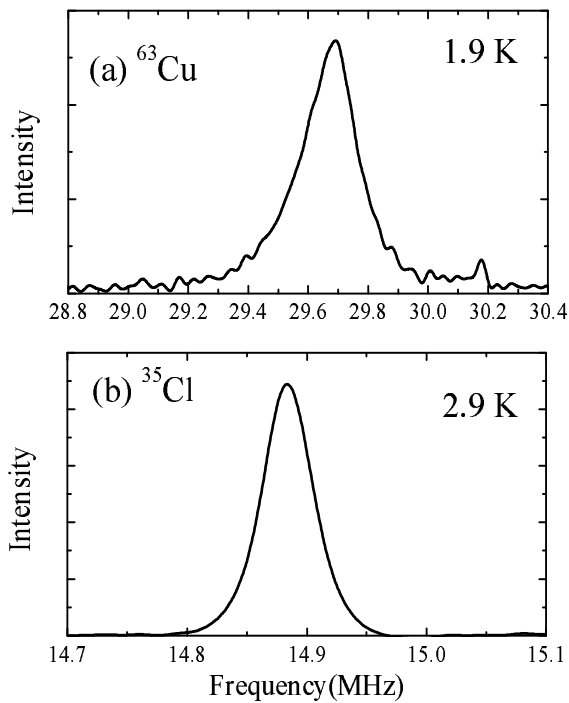}
\end{center}
\caption{NQR spectra (a) at the $^{63}$Cu sites ($T$ = 1.9 K) and (b) at the $^{35}$Cl sites ($T$ = 2.9 K).}
\label{nqr1}
\end{figure}  
Figure~\ref{nqr1} shows the NQR spectra at zero field for the $^{63}$Cu and $^{35}$Cl sites, 
both having the nuclear spin $I$ = 3/2. A sharp single NQR line is observed for each nuclei. 
This shows that both Cu and Cl atoms occupy a unique site and 
there is no substantial disorder for the Cu and Cl positions. The NQR frequencies $\nu _{NQR}$ 
are determined to be 29.68 MHz for $^{63}$Cu and 14.88 MHz for $^{35}$Cl.
The values of $\nu _{NQR}$ and the spectrum shape show no temperature dependence 
from 10 K to 1.5 K, indicating that the structure is unchanged 
at low temperatures. 

\begin{figure}[tb]
\begin{center}
\includegraphics[width=7.5cm]{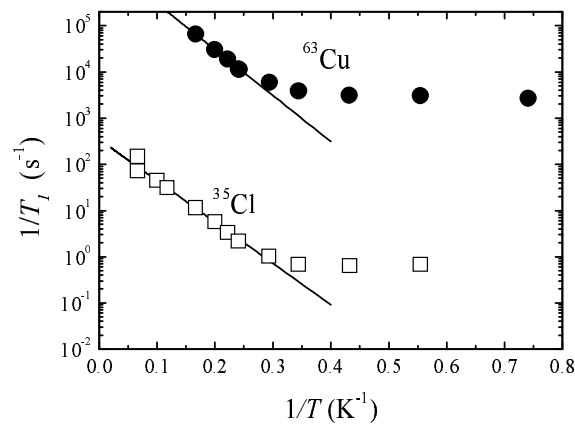}
\end{center}
\caption{Inverse temperature dependence of $1/T_1$ at the $^{63}$Cu and $^{35}$Cl sites. 
The solid lines show the fit to an activation law $1/T_1 \propto$ exp$(-E/k_{\mathrm{B}}T)$.}
\label{nqr2}
\end{figure}

Figure~\ref{nqr2} shows the inverse temperature dependence of $1/T_1$ at the $^{63}$Cu and $^{35}$Cl sites 
at zero field. At both sites, $1/T_1$ is well fitted by an activation law $1/T_1 \propto$ exp$(-E/k_{\mathrm{B}}T)$. 
The activation energy is obtained as $E/k_{\mathrm{B}}$ = 23 $\pm $ 2 K for Cu and $E/k_{\mathrm{B}}$ = 21 $\pm $ 1 K for Cl. 
Below 4 K, $1/T_1$ deviates from the activation law and become independent of temperature below 3 K. 
The origin of the constant relaxation rate at low temperatures is likely to be impurities. 

\subsection{Field swept NMR spectra and EFG tensors} 
\begin{figure}[tb]
\begin{center}
\includegraphics[width=7.5cm]{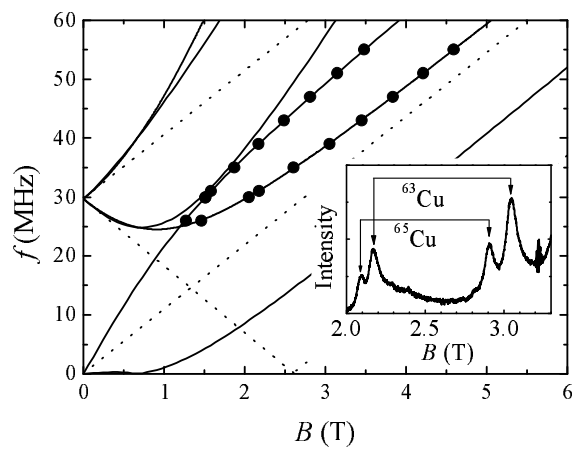}
\end{center}
\caption{Frequency-field diagram at 3 K at the $^{63}$Cu sites in the applied magnetic fields 
$B$ parallel to the $c$-axis. The dotted lines represent the resonance modes for the case of axially 
symmetric EFG around the $c$-axis. 
The solid lines are the fit to the results obtained from the Hamiltonian (2). 
The inset shows the Cu-NMR spectrum at 43 MHz. }
\label{efgcu}
\end{figure}

In the inset of Fig.~\ref{efgcu}, we show a typical field swept Cu-NMR spectrum 
for the aligned sample with the magnetic field $B$ applied parallel to the $c$-axis. 
The spectrum has four sharp peaks assigned to the $^{63}$Cu- and $^{65}$Cu-NMR 
lines from the aligned part of the sample. 
The spectrum also shows a broad tail spreading 
to higher field. This tail probably comes from the unaligned portion of the sample. 
The resonance fields are determined precisely from the peak positions. The frequency-field 
($f$-$B$) diagram can be constructed by plotting the resonance fields for 
various NMR frequencies between 26 and 55 MHz as shown in Fig.~\ref{efgcu}. 

The resonance frequency is generally determined from the following Hamiltonian 
including the magnetic hyperfine interaction and the electric quadrupole interaction,\cite{d2} 
\begin{equation}
\begin{split}
H = &-h\gamma (\boldsymbol{B} \cdot \boldsymbol{I} + 
\boldsymbol{I} \cdot \hat{K} \cdot \boldsymbol{B}) \\
&+ \frac{eQ}{6I(2I-1)} \sum_{\alpha , \beta }V_{\alpha \beta } 
\Big[ \frac{3}{2}(I_{\alpha } I_{\beta } + I_{\beta } I_{\alpha }) - \delta _{\alpha \beta }I^2 \Big], 
\end{split}
\end{equation} 
where $h$ is the Planck's constant, $Q$ is the nuclear quadrupole moment, 
$\gamma $ is the gyromagnetic ratio, $\hat{K}$ is the magnetic hyperfine shift tensor, 
$\delta _{\alpha \beta }$ is 0 (${\alpha \not= \beta }$) or 1 (${\alpha = \beta }$), 
and $V_{\alpha \beta } = \partial ^2V/\partial \alpha \partial \beta $ ($\alpha , \beta $ = $x$, $y$, or $z$)
is the component of the electric field gradient (EFG) tensor. 
If the Cu sites have the $C_4$-symmetry around the $c$-axis in the tetragonal 
crystal structure, $V_{\alpha \beta }$ is axially symmetric, that is, $V_{\alpha \beta }$ 
= 0 for ${\alpha \not= \beta }$ and $V_{zz} = - 2V_{xx}$ ($V_{xx} = V_{yy}$, $z \parallel c$). 
If the magnetic field is applied parallel to the symmetric $c$-axis, 
three resonance modes appear at the frequencies $\nu _R$ and $|\nu _{NQR} \pm \nu _R|$, 
where $\nu _R = (1 + K_{cc})\gamma B$ and $\nu _{NQR} = 3eQV_{zz}/2I(2I-1)h$, 
provided that all the components of $\hat{K}$ are much smaller than one. 
These results shown by the straight dotted lines in Fig.~\ref{efgcu} are apparently in 
contradiction with the nonlinear behavior of the experimental data. This gives the direct evidence that 
the Cu sites do not have axial symmetry around the $c$-axis. 

\begin{figure}[tb]
\begin{center}
\includegraphics[width=7.5cm]{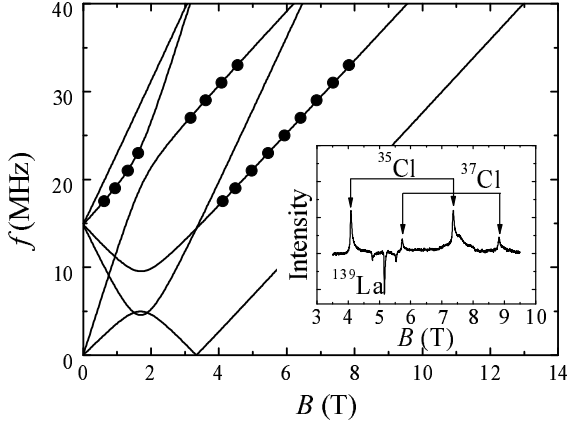}
\end{center}
\caption{Frequency-field diagram at 3 K at the $^{35}$Cl sites in the applied magnetic fields 
$B$ parallel to the $c$-axis. The solid lines show the fit obtained from the Hamiltonian (2). 
The inset shows the Cl-NMR spectrum at 31 MHz. }
\label{efgcl}
\end{figure} 

In the following, we assume that the $c$-axis is one of the principal axes of the EFG 
tensor at the Cu sites. This is the case, for example, when the Cu-Cl planes have the 
mirror symmetry. Although this assumption may not be strictly valid in the real structure, 
it is not possible to determine $V_{\alpha \beta }$ from the available data without any assumption. 
The nuclear Hamiltonian (1) is rewritten for the magnetic field parallel to the $c$-axis ($z$-axis) as 
\begin{equation}
\begin{split}
H = &-h\gamma B(1 + K)I_z \\
&+ \frac{h}{3}(\nu _xI_x^2 + \nu _yI_y^2 + \nu _zI_z^2), 
\end{split}
\end{equation} 
where $K$ is the magnetic hyperfine shift along the $c$-axis and 
$\nu _\alpha = \{3eQ/2I(2I-1)h \} V_{\alpha \alpha }$.\cite{d2} 
Although eq. (2) contains four parameters ($\nu _z$, $\nu _x$, $\nu _y$, and $K$), 
the number of independent parameters can be reduced to two ($K$ and $\nu _z$) from the 
relations $\nu _x + \nu _y + \nu _z = 0$ and $\nu _{NQR} = \sqrt{\nu _z^2 +(\nu _x - \nu _y)^2/3}$, 
where the NQR frequency $\nu _{NQR}$ is determined to be 29.68 MHz in section 3.1. 
Four eigenvalues of this Hamiltonian can be calculated 
analytically in the case of $I$ = 3/2, resulting in six resonance modes. 
As shown in Fig.~\ref{efgcu}, the experimental data 
are well fitted by the solid lines obtained from eq. (2) with $\nu _z$ = 16.26 $\pm $ 0.02 MHz, 
$\nu _x$ = 13.37 $\pm $ 0.02 MHz, $\nu _y$ = $-29.64$ $\pm $ 0.02 MHz, and $K$ = 0.44 $\pm $ 0.02 \%.

Figure~\ref{efgcl} shows the $f$-$B$ diagram at the $^{35}$Cl 
sites obtained from the field swept 
NMR spectra of the aligned sample. A typical spectrum is shown in the inset of Fig.~\ref{efgcl}. 
The applied magnetic field $B$ was parallel to the aligned direction ($c$-axis). 
The spectrum also shows the $^{139}$La-NMR lines. The negative intensity is due to improper 
rf-pulse conditions, which were optimized for Cl signal. 
The $f$-$B$ diagram can be fitted by solving 
the Hamiltonian (2) in the same way as has been done for the Cu sites. In Fig.~\ref{efgcl}, 
the solid lines show the results calculated for the parameter values $\nu _z$ = 14.17 $\pm $ 0.02 MHz, 
$\nu _x$ = $-$3.13 $\pm $ 0.02 MHz, $\nu _y$ = $-11.04$ $\pm $ 0.02 MHz, 
and $K$ = 0.04 $\pm $ 0.02 \%. 

\begin{figure}[tb]
\begin{center}
\includegraphics[width=7.5cm]{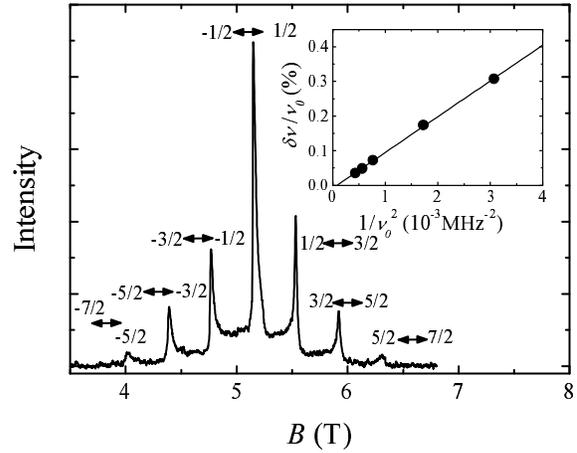}
\end{center}
\caption{Field swept NMR spectrum at the $^{139}$La sites at 6 K and 31 MHz 
in the magnetic field parallel to the $c$-axis. The inset shows the $\delta \nu/\nu _0$ versus 
$1/\nu _0^2$ plot at 115 K.}
\label{efgla}
\end{figure} 

Figure~\ref{efgla} shows the field swept $^{139}$La-NMR spectrum at 
6 K and 31 MHz in the magnetic 
field parallel to the $c$-axis. The seven peaks split by the quadrupole interaction 
for $I$ =7/2 are clearly observed. 
Since the quadrupole interaction at the La sites is much smaller than that at 
the Cu and Cl sites, the second order perturbation theory is sufficient to extract the values 
of $\nu _\alpha$ and $K$. We assume the $c$-axis to be one of the principal axes for EFG. 
The resonance frequency between the states $|m \rangle$ and $|m-1 \rangle$ is obtained 
from the Hamiltonian (2) as  
\begin{equation}
\begin{split}
\nu_{m \leftrightarrow m-1} = & \nu _0(1+K) - \Big(m-\frac{1}{2}\Big)\nu _z \\
&- \frac{(\nu _x - \nu _y)^2}{12\nu _0} \Big\{m(m-1)-\frac{19}{4}\Big\} 
\end{split}
\end{equation}
where $\nu _0$ = $\gamma B$.\cite{sp} If we define the frequency shift 
$\delta \nu _m \equiv  \nu _{m\leftrightarrow m-1} - \nu _0$, 
the frequency shift of the central line $\delta \nu _{1/2}$ obeys the relation, 
\begin{equation}
\begin{split}
\frac{\delta \nu_{1/2}}{\nu _0} = K + \frac{5}{12}\frac{(\nu _x - \nu _y)^2}{\nu _0^2}.
\end{split}
\end{equation}
Therefore, $\delta \nu_{1/2}/\nu _0$ plotted against $1/\nu _0^2$ should yield a straight line with 
the slope $(5/12)(\nu _x - \nu _y)^2$ and the intercept $K$ at $1/\nu _0^2$ = 0. 
The inset of Fig.~\ref{efgla} 
shows the $\delta \nu/\nu _0$ versus $1/\nu _0^2$ plot at 115 K. 
We find that $K$ at the La sites is almost zero ($-8 \times 10^{-3} \pm 2 \times 10^{-3}$ \%) and 
$|\nu _x - \nu _y|$ = 1.57 MHz. The value of $\nu _z$ is determined from the frequency shift for the 
satellite lines as $\nu _z = |\nu _{m\leftrightarrow m-1} - \nu _{-m+1 \leftrightarrow -m}|/(2m-1)$. 
The values of $\nu _z$, $\nu _x$, and $\nu _y$ are determined to be 2.23 $\pm $ 0.02, $-$0.33 $\pm $ 0.02, 
and $-$1.90 $\pm $ 0.02 MHz, respectively. 

\begin{table}[b]
\caption{Principal values of EFG tensors at the Cu, Cl, and La sites.}
\label{t1}
\begin{tabular}{cccc}
\hline
  & Cu & Cl & La \\
\cline{2-4}
$\nu _z$ (MHz) & 16.26 $\pm $ 0.02 & 14.17 $\pm $ 0.02 & 2.23 $\pm $ 0.02 \\
$\nu _x$ (MHz) & 13.37 $\pm $ 0.02 & $-$3.13 $\pm $ 0.02 & $-$0.33 $\pm $ 0.02 \\
$\nu _y$ (MHz) & $-$29.64 $\pm $ 0.02 & $-$11.04 $\pm $ 0.02 & $-$1.90 $\pm $ 0.02 \\
\hline
\end{tabular}
\end{table}

The values of $\nu _\alpha $ 
determined by NMR are listed in Table I for the Cu, Cl, and La sites. 
It should be noted that the values in Table I are determined on the 
assumption that the $c$-axis is one of the principal axes of EFG.
The EFG tensor is not axially symmetric around the $c$-axis for all sites. 
That is, none of these sites has the $C_4$-symmetry, which is incompatible with the reported 
tetragonal structure. Thus the $J_1$-$J_2$ model for the simple square 
lattice has to be modified for proper description of 
this system. We discuss modified models in section 4.3.

\subsection{Hyperfine shift and coupling constant} 
\begin{figure}[tb]
\begin{center}
\includegraphics[width=7.5cm]{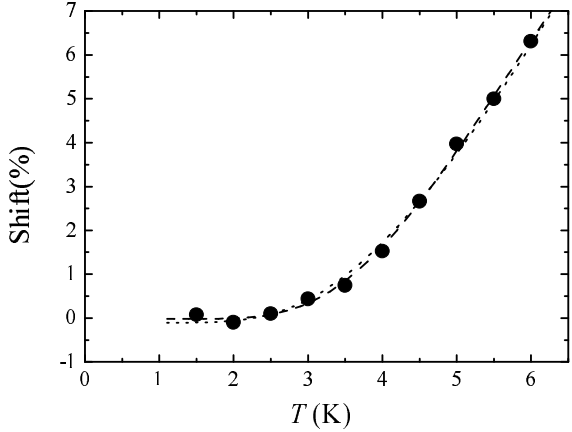}
\end{center}
\caption{Temperature dependence of $K$ at the $^{63}$Cu sites for the magnetic field parallel 
to the $c$-axis. The dashed and dotted lines show the calculated results for the isolated dimer model 
and the two dimensional coupled dimer model, respectively.}
\label{shift}
\end{figure} 
By repeating the above analysis for different temperatures, we obtained the 
temperature dependence of $K$. 
Figure~\ref{shift} shows the temperature dependence of 
$K$ at the Cu sites in the magnetic field parallel to the $c$-axis. 
The shift consists of the contributions from 
the spin and Van Vleck susceptibility at the Cu sites, $K$ = $K_s + K_{\mathrm{VV}}$. 
Each term is related to the spin and the Van Vleck susceptibility 
as $K_s = (A_{hf}/N_\mathrm{A}\mu_{\mathrm{B}})\chi _s$, $K_{\mathrm{VV}} = 
(A_{orb}/N_{\mathrm{A}}\mu_{\mathrm{B}})\chi _{\mathrm{VV}}$. 
The chemical shift is expected to be negligibly small compared with the 
spin and orbital shifts due to $d$-electrons of Cu$^{2+}$. 
As shown in Fig.~\ref{shift}, $K$ decreases with decreasing $T$ and is 
extrapolated to zero at $T$ = 0. This means $K_s$ and $A_{hf}$ are positive. 
Since $K_{\mathrm{VV}}$ is always positive, we can conclude that $K_{\mathrm{VV}}$ is nearly zero 
and $K_s$ approaches zero at $T$ = 0. It should be noted that while $\chi $ contains 
impurity contribution, $K$ is free from impurities, and thus provides direct evidence for a spin-gap. 

We estimate the spin-gap in two different models. 
The dashed line in Fig.~\ref{shift} is the fit to the isolated dimer model 
\begin{equation}
K=K_{\mathrm{VV}} + \frac{C}{T}\Big[1+\frac{1}{3}\mathrm{exp}(E'/k_{\mathrm{B}}T) \Big]^{-1}, 
\end{equation}
from which we obtained $E'/k_{\mathrm{B}}$ = 21.8 K and $K_{\mathrm{VV}} = -0.02 \pm 0.02$ \%. 
Note that the shift was determined in the range of magnetic field from 1 to 5 T. 
Considering the reduction of the gap by the average field 
$g\mu _{\mathrm{B}} \langle B \rangle /k_{\mathrm{B}} \simeq $ 4 K ($\langle B \rangle$ = 3T), 
the gap at zero field is estimated to be 26 K. This value is rather close to the gap 
($\sim $27 K) estimated from the neutron inelastic scattering and the susceptibility measurements.\cite{b1,b2} 
However, the isolated dimer model ignores the dispersion of the triplet excitations. 
In the opposite limit, where the triplets have large dispersion $\epsilon_q = E'' + cq^2$, 
the spin susceptibility is determined by the excitation near the bottom of the dispersion 
$K_s \propto (1/T) \int D(\epsilon) n(\epsilon) d\epsilon \propto \mathrm{exp}(-E''/K_{\mathrm{B}}T)$, 
where $D(\epsilon)$ is the density of state, which is constant for $\epsilon \geq E''$ in two dimension, 
and $n(\epsilon)$ is the Bose factor. The dotted line in Fig.~\ref{shift} shows the fit to this 
model with $E''/k_{\mathrm{B}}$ = 15 K. Therefore, the gap is estimated to be 19 K at zero field. 
The gap values obtained in this study are compared with the reported results in section 4.1.

\begin{figure}[tb]
\begin{center}
\includegraphics[width=7.5cm]{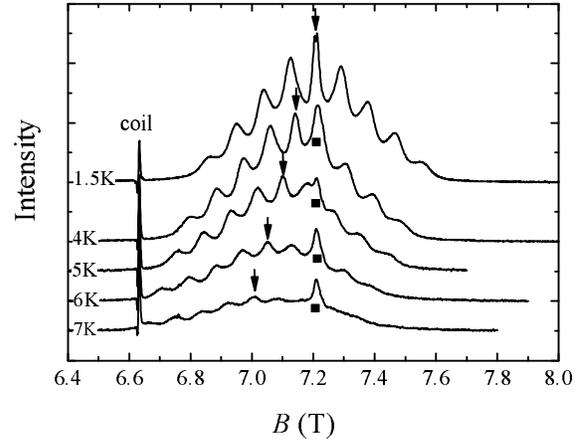}
\end{center}
\caption{Temperature dependence of the field swept spectrum at the Nb sites at 75 MHz
for the applied magnetic fields $B$ parallel to the $c$-axis.}
\label{tdnb}
\end{figure} 

Figure~\ref{tdnb} shows the temperature dependence of the field swept 
NMR spectrum at the Nb sites. 
The nine peaks for $I$ = 9/2 is observed at 1.5 K and the central line is indicated by the arrow. 
The central and the eight satellite lines shift to lower fields 
with increasing temperature, indicating sizable hyperfine coupling at the Nb sites 
to spin magnetization of Cu ions. 
In addition, we observed another peak marked by the solid square 
in Fig.~\ref{tdnb}. This peak does not shift with temperature, indicating no hyperfine 
coupling to the Cu spins. In fact, $1/T_1$ for this peak is ten times smaller than that 
for the intrinsic signal at low temperatures. Therefore, we conclude that 
this signal comes from extrinsic nonmagnetic phases. 
Imperfection of the ion-exchange reaction is a possible origin of such 
nonmagnetic impurely phases. 

\begin{table}[b]
\caption{Hyperfine coupling constants for $B \parallel c$.}
\label{t2}
\begin{tabular}{cccc}
\hline
  & Cu & Cl & Nb \\
\cline{2-4}
$A_{hf}$ (T/$\mu _{\mathrm{B}}$) & 13.8 $\pm $ 0.3 & 6.9 $\pm $ 0.5 & 2.84 $\pm $ 0.08 \\
\hline
\end{tabular}
\end{table} 

\begin{figure}[tb]
\begin{center}
\includegraphics[width=7.5cm]{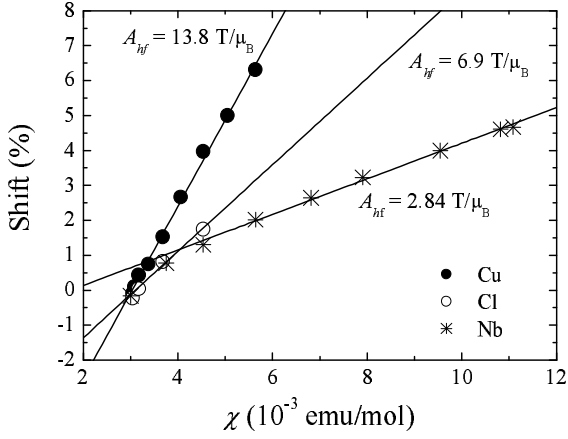}
\end{center}
\caption{$K$-$\chi$ plot for the Cu, Cl, and Nb sites.}
\label{kkai}
\end{figure}

Figure~\ref{kkai} shows the $K$-$\chi$ plot for various sites obtained for the temperature range 1.5 $\sim $ 20 K. 
Here, a Curie term due to magnetic impurities was removed from the $\chi $ data as described in 
section 3.1. The hyperfine coupling constants $A_{hf}$ for $B \parallel c$ are determined to be 
13.8 $\pm $ 0.3, 6.9 $\pm $ 0.5, and 2.84 $\pm $ 0.08 T/$\mu _{\mathrm{B}}$ for the Cu, Cl, and Nb sites, respectively, 
(Table II). 
Although the NMR shifts at all sites approach near zero as $T \rightarrow 0$, $\chi $ maintains 
a finite value ($3 \times 10^{-3}$ emu/mol) as mentioned in section 3.1. Note that the impurity 
term from free spins has been already subtracted. The residual $\chi $ at low temperatures, which corresponds to 
the $B$-linear magnetization in the inset of Fig.~\ref{kai}, does not couple to any nuclei observed by NMR. 
Hence we conclude that it has as extrinsic origin. 

\subsection{High magnetic field region} 
\begin{figure}[tb]
\begin{center}
\includegraphics[width=7.5cm]{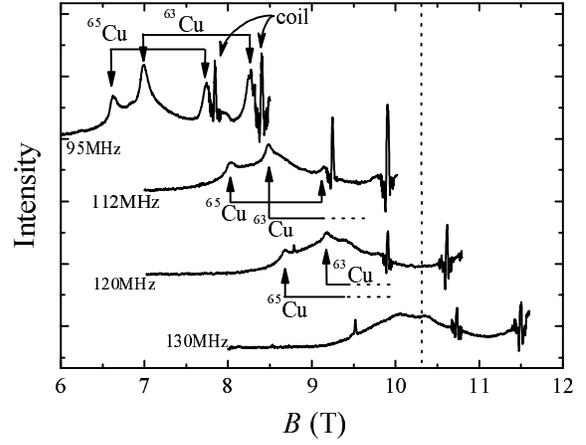}
\end{center}
\caption{Frequency dependence of the field swept NMR spectrum at the Cu sites at 1.5 K 
in the applied magnetic fields $B$ parallel to the $c$-axis. 
The dotted lines show the critical field $B_{c1}$.}
\label{hfcu}
\end{figure} 

Figure~\ref{hfcu} shows the field swept NMR spectrum at 1.5 K for the Cu sites at different frequencies in 
the higher magnetic field region above $B_{c1} \simeq  $ 10.3 T. The very sharp peaks are due 
to the NMR-coil made from Cu. The $^{63,65}$Cu-NMR lines from the aligned sample are indicated by the arrows. 
Their intensity decreases with increasing $B$ and the peaks from the aligned sample disappear 
slightly below $B_{c1}$ =10.3 T. This is because the spin-spin relaxation 
time $T_2$ is getting shorter when the field approaches $B_{c1}$. There remains a broad signal near $B_{c1}$, 
observed in the spectrum at 130 MHz. Since this signal has much longer $T_2$, it is probably due to some impurity phases. 

\begin{figure}[t]
\begin{center}
\includegraphics[width=7.5cm]{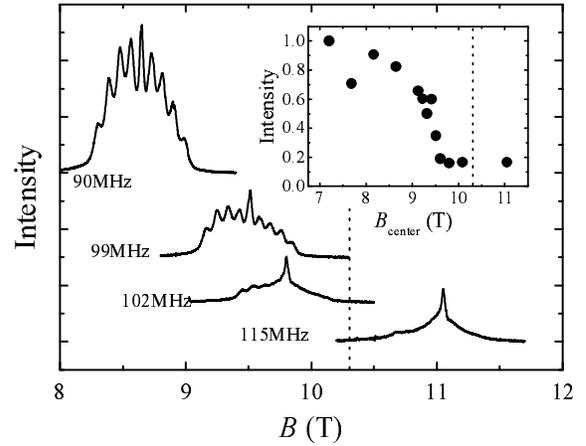}
\end{center}
\caption{Frequency dependence of the field swept NMR spectrum at the Nb sites at 1.5 K 
in the applied magnetic fields $B$ parallel to the $c$-axis. 
The inset shows the $B_{center}$-dependence of the integrate intensity. 
The dotted lines show the critical field $B_{c1}$.}
\label{hfnb}
\end{figure} 

Figure~\ref{hfnb} shows the frequency dependence of the field swept 
spectrum at the Nb sites. 
The quadruple split peaks for $^{93}$Nb ($I$ = 9/2) are observed at 90 MHz. 
The intensity of the split peaks decreases with increasing $B$ and the signal disappears slightly below 
$B_{c1}$ in a similar way as has been observed for the Cu signals. 
Above $B_{c1}$, the spectrum with one central peak and broadened quadrupole satellites 
is observed (115 MHz). This signal probably has the same origin as the unshifted signal 
observed at 75 MHz at high temperatures (Fig.~\ref{tdnb}). In the inset of Fig.~\ref{hfnb}, 
the integrated intensity of the whole spectrum is plotted against the field of the central line ($B_{center}$). 
The intensity normalized by the value at 7.2 T decreases with increasing 
$B_{center}$ and vanishes at 9.5 T, slightly below $B_{c1}$, 
except for the contribution from impurity phases, indicating that a field-induced magnetic phase 
transition occurs around $B_{c1}$. The field-induced transition is also indicated by the recent specific 
heat measurements in magnetic fields.\cite{sf}

\subsection{TEM result} 
\begin{figure}[t]
\begin{center}
\includegraphics[width=7cm]{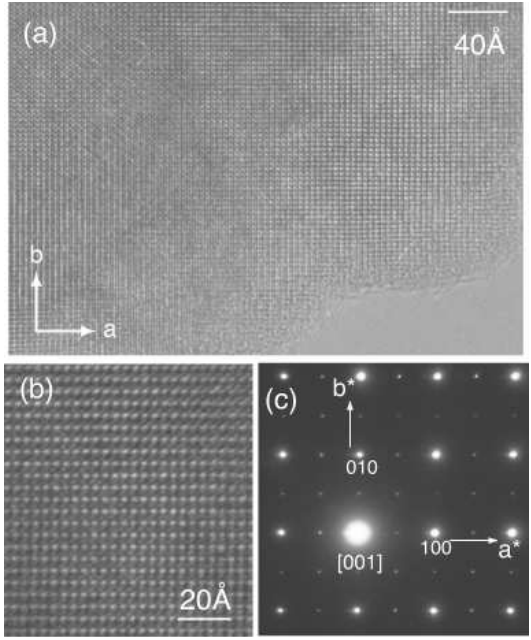}
\end{center}
\caption{(a) High-resolution TEM image of (CuCl)LaNb$_2$O$_7$ at room temperature 
along the [001]-zone 
axis and (b) its enlarged view. (c) Corresponding electron 
diffraction pattern, where indices 
are given based on the tetragonal unit cell of the $P4/mmm$ structure.\cite{c1}}
\label{tem}
\end{figure}

In order to obtain further information about the crystal structure, we performed TEM measurements 
at room temperature. Shown in Fig.~\ref{tem} (a) and (b) are the [001] zone lattice images that demonstrate 
contrasts with a special period of about 4 {\AA} along $\langle 100\rangle $, consistent with the 
lattice parameter. However, the corresponding electron diffraction (Fig.~\ref{tem} (c)) has revealed 
weak reflections such as (1/2 0 0) or (0 1/2 0), 
in addition to strong fundamental reflections relevant to the reported crystal structure. 
The weak reflections are commensurate and indicate 
the doubling of the lattice period of both $a$- and $b$-axes, i.e., the size of the unit cell is 
given by $2a \times  2b$. 
Such a superstructure has not been detected even by high $q$-resolution synchrotron X-ray diffraction, 
\cite{tem} presumably due to tiny modulation of the atoms and/or small domain size. 
Note that the superstructure reflection at (1/2 1/2 0) was observed in (FeCl)LaNb$_2$O$_7$.\cite{tem}

\section{Discussion} 

\subsection{Spin-gap} 
The gap value in (CuCl)LaNb$_2$O$_7$ is still controversial, while 
it has been investigated by several experimental methods. The most 
direct method is the inelastic neutron scattering, which shows the 
magnetic excitation with the gap energy of 2.3 meV (26.7 K).\cite{b1} 
Although it is possible that the limited energy resolution masks fine 
structure of the neutron spectrum and the true gap may be somewhat smaller, 
the large discrepancy between this value and the gap derived from the 
critical magnetic field, $g\mu_{\mathrm{B}} B_{c1}/k_{\mathrm{B}}$ = 15 K, cannot be explained by such an effect. 

From the susceptibility measurements, the gap is estimated to 
be 27 K by using the isolated dimer model.\cite{b1} The magnetic 
hyperfine shift determined by NMR is free from impurity contribution and 
provides more reliable spin susceptibility. 
Using the isolated dimer model, the gap value is determined to be 26 K, 
consistent with the susceptibility results.  
We have also examined a two-dimensional coupled dimer model with a large width of the triplet 
dispersion and obtained the gap value of 19 K. 
Since these two models correspond to the opposite limits of zero and infinite width 
of the dispersion, the values 26 K and 19 K set the upper and the lower bounds for the true gap. 
From the NQR relaxation rate $1/T_1$ at the 
Cu and Cl sites, the gap value is determined to be 23 $\pm $ 2 and 21 $\pm $ 1 K, 
respectively, consistent with the above range. 
This range is still significantly larger than $g\mu_{\mathrm{B}} B_{c1}/k_{\mathrm{B}}$ (= 15 K). 

To reconcile the discrepancy between the gap estimated at low magnetic fields and $B_{c1}$, 
Kageyama {\it et al}. proposed formation of two-triplet bound states with $S$ = 2.\cite{b2} 
If two-triplet bound states are stabilized due to frustration, the gap can close before  
$B$ reaches $\Delta/g\mu_{\mathrm{B}}$ and a nematic order may be realized above $B_{c1}$.\cite{n1} 
The Cu- and Nb-NMR signals were undetectable above $B_{c1}$ at 1.5 K due to divergence of $1/T_2 $ 
as mentioned before (Figs.~\ref{hfcu} and~\ref{hfnb}). 
To investigate the magnetic structure above $B_{c1}$, we need measurements 
at higher fields and lower temperatures.

\subsection{$d$-orbital} 

\begin{figure}[tb]
\begin{center}
\includegraphics[width=7.5cm]{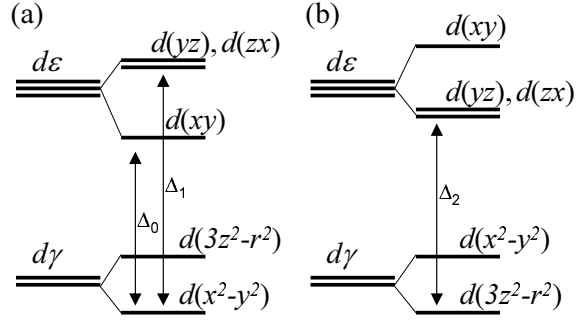}
\end{center}
\caption{Orbital energy levels for Cu$^{2+}$ in a tetragonal crystal field. 
The ground state wave function is $d(x^2 - y^2)$ in (a) or $d(3z^2 - r^2)$ in (b).}
\label{energy}
\end{figure} 

The experimental determination of $A_{hf}$ and $K_{\mathrm{VV}}$ at the Cu sites provides information about 
the Cu $d$-orbital. In this section, we compare the experimental results of $A_{hf}$ and $K_{\mathrm{VV}}$ with 
the calculated values. Figure~\ref{energy} shows the orbital 
energy levels for Cu$^{2+}$ in tetragonal crystal fields. 
The Cu$^{2+}$ ions have one $d$-hole with either $d(x^2 - y^2)$ or $d(3z^2 - r^2)$ wave function. 
The hyperfine coupling constants $A_\parallel ^{hf}$ (for $B \parallel  z$) and $A_\perp ^{hf}$ 
(for $B \perp z$) are expressed as 
\begin{equation}
\begin{split}
A_{\parallel}^{hf} &= \frac{\mu _0 \mu _{\mathrm{B}}}{4\pi } \langle r^{-3} \rangle \Big( -\kappa - \frac{4}{7}
 +  \frac{6}{7} \frac{\lambda }{\Delta _1} -  \frac{8\lambda }{\Delta _0} \Big) \\
A_{\perp} ^{hf} &= \frac{\mu _0 \mu _{\mathrm{B}}}{4\pi } \langle r^{-3} \rangle \Big( -\kappa + \frac{2}{7}
 +  \frac{11}{7} \frac{\lambda }{\Delta _1} \Big) \ \ \mathrm{for} \  d(x^2 - y^2) \\
A_{\parallel} ^{hf} &= \frac{\mu _0 \mu _{\mathrm{B}}}{4\pi } \langle r^{-3} \rangle \Big( -\kappa + \frac{4}{7} 
+  \frac{6}{7} \frac{\lambda }{\Delta _2} \Big) \\
A_{\perp} ^{hf} &= \frac{\mu _0 \mu _{\mathrm{B}}}{4\pi } \langle r^{-3} \rangle \Big( -\kappa - \frac{2}{7} 
-  \frac{45}{7} \frac{\lambda }{\Delta _2}  \Big) \ \ \mathrm{for} \ d(3z^2 - r^2) , 
\end{split}
\end{equation}
where $\mu _0$ is the permeability of vacuum, $\kappa $ is the core polarization 
coefficient, $\lambda $ is the spin-orbit coupling parameter, and $\Delta _i$ ($i$ = 0, 1, 2) is 
the energy between the ground state and the excited sate shown in 
Fig.~\ref{energy}.\cite{e1,e2} The values of $\kappa $, $\lambda $, and $\langle r^{-3} \rangle$ 
for Cu$^{2+}$ ions are common to all insulating materials within 10-20 \%, typically given as 
$\kappa $ = 0.28, $\lambda $ = $-8.8 \times  10^{-2}$ eV, 
and $\langle r^{-3} \rangle$ = $4.26 \times 10^{31}$ m$^{-3}$.\cite{e1,e2} 
$\Delta _i$ depends on materials but should be of the order of a few eV. 
We take $\Delta _i$ = 2.0 eV.\cite{e1,e2} 
By using these values, $A_\parallel ^{hf}$ and 
$A_\perp ^{hf}$ are estimated to be $-$24.1 and 2.71 T/$\mu _{\mathrm{B}}$ for $d(x^2 - y^2)$ and 
10.2 and $-$12.2 T/$\mu _{\mathrm{B}}$ for $d(3z^2 - r^2)$, respectively. 
Compared with the experimental values, $A_{hf}$ (=13.8 T/$\mu _{\mathrm{B}}$), 
$A_\parallel ^{hf}$ for $d(x^2 - y^2)$ and $A_\perp ^{hf}$ for $d(3z^2 - r^2)$ can be excluded 
because they lead to large negative shifts. 
Note that the magnetic field is parallel to the $c$-axis in the experiment, and we consider the cases 
where the $z$-axis of the orbital is either parallel or perpendicular to the $c$-axis. 
Among the other two cases, $A_\parallel ^{hf}$ for $d(3z^2 - r^2)$ is by far the favorable.

The vanishing Van Vleck susceptibility $K_{\mathrm{VV}} \simeq  $  0 for $B \parallel c$ 
also supports the $d(3z^2 - r^2)$ orbital. 
For $B \parallel z$, the Van Vleck susceptibility $\chi _{\mathrm{VV}}$ (emu/mol) is calculated as 
\begin{equation}
\chi _{\mathrm{VV}} = N_{\mathrm{A}} \mu _{\mathrm{B}}^2 
\sum_{n \not= 0} \frac{|\langle \psi_n |L_z|\psi _0 \rangle |^2}{\epsilon _n - \epsilon _0},
\end{equation}
where $\psi _0$ is the grand state with the energy $\epsilon _0$ and $\psi _n$ is the excited 
state with the energy $\epsilon _n$. For the $d(3z^2 - r^2)$ orbital, $\chi _{\mathrm{VV}}$ is zero. 
On the other hand, $\chi _{\mathrm{VV}} $ for $d(z^2 - x^2)$ is calculated to be 
$N_{\mathrm{A}} \mu _{\mathrm{B}}^2/\Delta _1$ $\simeq $ $1.6 \times 10^{-5}$ emu/mol, leading to 
$K_{\mathrm{VV}} = 2\langle r^{-3} \rangle \chi _{\mathrm{VV}}/N_{\mathrm{A}}$ = 0.23 \%. 
This is clearly incompatible with the experimental observation.  
Therefore, we conclude that the Cu spins are on the $d$-orbital 
consisting mainly of $d(3z^2 - r^2)$. This is natural because the Cu-O bonding (1.84 \AA) 
is much shorter than the Cu-Cl bonding in the $ab$ plane (3.14 or 2.40 \AA).\cite{c2} 
Such structure and the difference of the ionic valence (O$^{2-}$ vs. Cl$^-$) make the $d(3z^2 - r^2)$ 
orbital favorable for the $d$-hole. 
It should be noted that broken axial symmetry revealed by our NMR data should allow 
slight mixing of other components such as $d(x^2 - y^2)$. 
The large hyperfine field at the Nb sites can then be understood as due to the transferred hyperfine 
interaction through the $d(3z^2 - r^2)$ orbital. We suspect that  
the interplane exchange interaction through the path Cu-O-Nb-O-Nb-O-Cu and the nearest-neighbor 
exchange interaction through the path Cu-O-O-O-Cu may not be completely negligible. 

\subsection{Possible model} 
Precise structural information of the CuCl plane is crucial to understand the origin of 
the spin-gap in this compound. Our NQR results show that both Cu and Cl have 
only one site and there is no apparent disorder. This means that periodic superstructure 
should be realized, even though the domain size of the superstructure may be very small. 
Although Caruntu {\it et al.} proposed disorder in the Cl position based 
on their neutron diffraction experiments,\cite{c2} 
their results are not necessarily incompatible with the NQR results because 
possibility of a superstructure was not considered in the analysis of the neutron data. 
In addition to the NQR results, the EFG tensor does not have axial symmetry at the Cu, Cl, and La sites and 
the TEM result shows that the unit cell is doubled along both the $a$- and the $b$-axes from the original 
square lattice. Therefore, $J_1$-$J_2$ model for the simple square lattice has to be modified. 
Whangbo and Dai proposed models of ring clusters with even number of spins based on the neutron results 
to account for the spin-gap. They lead, however, to several inequivalent Cu and Cl sites, 
inconsistent with our NQR results. 

\begin{figure}[tb]
\begin{center}
\includegraphics[width=7.5cm]{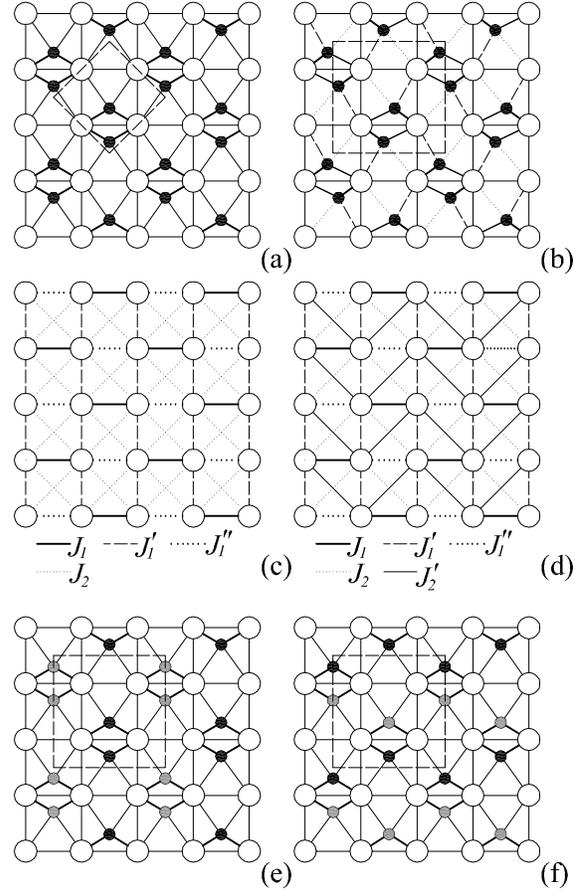}
\end{center}
\caption{Possible structures for the CuCl plane. The open and solid circles 
represent the Cu and Cl ions, respectively. The dashed lines in (a), (b), (e), and (f) 
indicate the unit cell for the superstructure.
In the structure (e) and (f), the solid and gray spheres represent the 
Cl atoms above and below the Cu plane, respectively.}
\label{model}
\end{figure} 

The rather large spin-gap suggests that the real structure has such a feature that allows natural 
formation of Cu dimers. Figures~\ref{model} (a), (b), (e), and (f) show examples of 
such structures, where we assume that Cu maintains the ideal position for the square 
lattice, (1/2, 1/2, 1/2). The structure in Fig.~\ref{model} (a) is simplest and leads to 
a two-dimensional dimer system as shown in Fig.~\ref{model} (c). Here, dimers are
formed by the dominant exchange path $J_1$, and connected 
with each other by $J_1'$, $J_1''$, and $J_2$. If $J_1$ is 
antiferromagnetic, a spin-gap is naturally formed. This structure has 
only one Cu and Cl site. 
The neutron diffraction study suggests that Cl moves off the ideal (0, 0, 1/2) 
position to one of the ($x$, 0, 1/2), ($-x$, 0, 1/2), (0, $x$, 1/2), and (0, $-x$, 1/2) positions 
in a random manner.\cite{c2} 
The structure in (a) is an ordered version of the structure 
proposed from the neutron experiments.\cite{c2} 
However, the unit vectors $\hat{a} \pm \hat{b}$ of the structure (a) leads to the 
superstructure reflection at (1/2 1/2 0) but 
cannot account for the observed TEM reflections such as (1/2 0 0) and (0 1/2 0). 
Therefore, Cl should take more general positions. We propose three models 
(b), (e), and (f), which are consistent with both the NQR and TEM results. In the structure 
(b), Cl moves off the ideal (0, 0, 1/2) position to a position, 
($x$, $y$, 1/2), ($-x$, $y$, 1/2), ($x$, $-y$, 1/2), or ($-x$, $-y$, 1/2), 
still maintaining only one site. The dominant exchange path $J_1$ makes the dimers, which 
are connected by other exchange paths 
$J_1'$, $J_1''$, $J_2$, and $J_2'$ as shown in Fig.~\ref{model} (d). In the 
structures (e) and (f), Cl moves off to 
($x$, 0, 1/2$+\delta $), ($x$, 0, 1/2$-\delta $), ($-x$, 0, 1/2$+\delta $), or ($-x$, 0, 1/2$-\delta $). 
The solid and gray spheres 
represent the Cl atoms above and below the Cu plane, respectively. 
In both (e) and (f), the exchange paths are the same as the structure 
(a) that is shown in Fig.~\ref{model} (c). Both models (c) and (d) generate 
frustration. 

A useful insight is obtained by comparing the magnetic ground state of 
(CuCl)LaNb$_2$O$_7$ and (CuBr)LaNb$_2$O$_7$, which shows a collinear antiferromagnetic order.\cite{b3} 
We have observed a large internal field of 16 T at the $^{79,81}$Br sites in (CuBr)LaNb$_2$O$_7$ 
by zero field NMR. The details will be published in a separate paper. 
If $J_1$ is antiferromagnetic in (CuBr)LaNb$_2$O$_7$, the internal field at the Br sites 
should be canceled out for the structures (a), (e), and (f). Therefore, if the 
two materials have the same structure, (b) is the only possibility. 
However, the exchange interaction through Cl or Br is very sensitive to small structural variation 
and often changes from ferromagnetic to antiferromagnetic.\cite{f3} Therefore, $J_1$ for 
(CuBr)LaNb$_2$O$_7$ may be ferromagnetic, even if $J_1$ for 
(CuCl)LaNb$_2$O$_7$ is antiferromagnetic. If $J_1$ is ferromagnetic 
in the structures (a), (e), and (f), a large internal field of 
the ordered moment survives. Therefore, both models (c) and (d) are possible. 
It seems that these examples exhaust possible structures within the condition of the 
uniqueness of the Cu and Cl sites and the doubled periodicity along the $a$- and $b$-axes, although 
we are not able to prove rigorously. 

\section{Summary} 
NMR, NQR, and TEM results of (CuCl)LaNb$_2$O$_7$ have been 
reported. A sharp single NQR line observed at each Cu and Cl 
site indicates that both Cu and Cl atoms occupy a unique site. 
The electric field gradient tensor shows that there is no 
axial symmetry around the $c$-axis at the Cu, Cl, and La sites. These results indicate 
that the $J_1$-$J_2$ model for the simple square lattice has 
to be modified. We propose a two-dimensional dimer model for 
this compound. Cu- and Nb-NMR signals disappear above the 
critical field $B_{c1} \simeq  $ 10 T at low temperatures, 
indicating a field-induced magnetic phase transition occurs at $B_{c1}$. 

\section*{Acknowledgment} 
We would like to thank T. Momoi, N. Shannon, H. Nojiri and K.-Y. Choi for valuable discussions. 
This work was supported by a Grant-in-Aid for Scientific Research (Nos. 18740202 and 17684018) 
and a Grant-in-Aid on Priority Areas (``Invention of Anomalous Quantum Materials'' 
Nos. 16076204 and 16076210) from the Ministry of Education, Culture, Sports, Science 
and Technology of Japan.

\end{document}